\documentstyle[preprint,aps]{revtex}

\tightenlines

\def\be{\begin{equation}}
\def\ee{\end{equation}}
\def\bea{\begin{eqnarray}}
\def\eea{\end{eqnarray}}
\def\ba{\begin{array}}
\def\ea{\end{array}}
\def\psibar{\bar{\psi}}

\def\lambdabar{\bar{\lambda}}
\def\Ap2{A(p^2)}
\def\Aq2{A(q^2)}
\def\Bp2{B(p^2)}
\def\Bq2{B(q^2)}
\def\slash{\gamma \cdot}
\def\half{\frac{1}{2}}
\def\Tr{\mbox{Tr}}
\def\inverse{\frac{1}}

\begin{document}
\setlength{\unitlength}{1mm}

\title{SUPERSYMMETRY AND CHIRAL SYMMETRY}

\author{M. L. WALKER\footnote{E-mail:mlw105@rsphysse.anu.edu.au}}

\address{Department of Theoretical Physics,\\ Research School of 
Physical Sciences and 
Engineering,\\ Australian National University, Canberra, A.C.T. 0200
Australia}

\maketitle

\begin{abstract}
We dispute a recent claim for a nonperturbative
nonrenormalisation theorem stating that mass cannot be spontaneously generated
in supersymmetric QED. We also extend a long-standing perturbative result,
namely that the effective potential is zero to all orders of perturbation
theory, to the nonperturbative regime for arbitrary numbers of flavours.
\end{abstract}

\setcounter{equation}{0}
\section{INTRODUCTION}
The supersymmetric (SUSY) 
nonrenormalisation theorem for mass to all orders in perturbation 
theory has been established for some time~\cite{WZ74a,IZ74,W90}.
Several authors~\cite{CL88,KS97,CS99a} have investigated the possibility of 
a nonperturbative nonrenormalisation theorem in SUSY quantum electrodynamics
(SQED), which asserts that the 
chiral solution to the Dyson-Schwinger equation (DSE) 
in SQED is not merely favoured but unique, 
ie.~that there is no achiral solution. 
The first was Clark and Love \cite{CL88} who, using the superfield formalism
and after truncating diagrams containing seagull and higher order $n$-point 
vertices, found that the effective mass $\cal M$ contains a prefactor $\xi - 1$
which vanishes in Feynman gauge. Reasoning that if the mass vanishes in
one gauge then it must vanish in all gauges, they conclude that there can be 
no dynamic chiral symmetry breaking in SQED, even beyond the rainbow 
approximation.

This approach was criticized by 
Kaiser and Selipsky on two grounds~\cite{KS97}. Firstly they argue that 
the truncation of seagull diagrams is too severe as it ignores contributions 
even at the one-loop level.  Secondly they point out that infinities 
arising from the infrared divergences which plague the superfield formalism 
can counter the vanishing prefactor and allow spontaneous mass generation. 
They also point out that the original nonrenormalisation theorem did not forbid
mass corrections, only infinite mass counter-terms.

The issue was taken up by 
Campbell-Smith and Mavromatos~\cite{CS99a} who investigated chiral
symmetry breaking in $2+1$ dimensional 
SQED (SQED$_3$) using superfields with both two- and 
four-component spinors. In the four-component theory~\cite{CS99a} 
they also find a nonrenormalisation theorem. Their analysis dimensionally
reduces SQED$_4$ to SQED$_3$, introducing a compactification scale in the
process. After truncating all two-particle irreducible diagrams from the DSE,
taking the limit that all momenta are small compared to the momentum scale 
of the compactification, 
and making several other approximations, they find the same prefactor in front 
of the effective mass as Clark and Love, and claim that its cancellation
by infrared divergences is subverted by the lack of a corresponding prefactor
in the renormalisation factor $\cal Z$. Since their argument depends on 
dimensional reduction of SQED$_4$, it cannot be applied in $3+1$ dimensions.

Our analysis in the component formalism finds no evidence for such a theorem.
It is certainly the case that no vanishing gauge dependent prefactor emerges.

The CJT effective potential~\cite{CJT74} expresses the potential of a theory
as a functional of its propagators and can be used to compare the chiral and
achiral propagators in ordinary field theories. It has been
known for some time that in perturbation theory~\cite{Z75,W76} the effective 
potential is exactly zero to all orders in a SUSY theory. (Pisarski has
adapted these proofs to the many flavour limit in the nonperturbative
theory~\cite{P84}.) However, a rigorous result in perturbation
theory cannot be assumed to hold in the nonperturbative regime. 
So while the favoured solution must have a potential
of exactly zero since SUSY is unbroken~\cite{W82}, 
it is still reasonable to ask if the unfavoured one does not.

There has been some confusion regarding the calculation of the CJT effective
potential (for example see~\cite{WB99a}) in SUSY theories. 
This arises from the fact that it is 
not clear how to treat Green's functions with auxiliary fields. This problem
was solved in a previous paper~\cite{WB99b}, and substituting the propagators
from that paper into the CJT effective potential yields that it is uniformly
zero. This renders the CJT effective potential ineffective for choosing between
solutions to the DSE in the absence of SUSY breaking.

Sec.~\ref{sec:theorem} gives our argument against the existence of a 
nonperturbative nonrenormalisation theorem. Substituting in the most general
possible form for the three-point vertices~\cite{WB99b}, we find that it is 
impossible to find an acceptable {\em ansatz} in which the effective mass must 
be zero. However we do not succeed in finding an achiral solution so the
(remote) possibility that the DSE cannot be solved with a dressed mass 
remains open.

Our analysis of the CJT effective potential, in which we show that it is
uniformly zero, is presented in Sec.~\ref{sec:CJT}. This result is not spoiled 
by vacuum polarisation.

\setcounter{equation}{0} 
\section{CHIRAL SYMMETRY BREAKING IS PERMITTED IN SQED}
\label{sec:theorem}

The Lagrangian of SQED is
\bea \label{lagrang}
L &=& |f|^2 + |g|^2+ |\partial_\mu a|^2 + |\partial_\mu b|^2
- \bar{\psi} \not \! \partial \psi \nonumber \\ \nonumber \\
&&-m(a^*f + af^* + b^*g + bg^* + i\bar{\psi} \psi)\nonumber \\ \nonumber \\
&&- ieA^\mu(a^\ast \stackrel{\leftrightarrow}{\partial}_\mu a
+ b^\ast \stackrel{\leftrightarrow}{\partial}_\mu b
+ \bar{\psi} \gamma_\mu \psi) \nonumber \\ \nonumber \\
&&- e[\bar{\lambda}(a^\ast + i\gamma_5 b^\ast)\psi 
- \bar{\psi}(a + i\gamma_5 b) \lambda] \nonumber \\ \nonumber \\
&&+ ieD(a^\ast b - a b^\ast)
+e^2 A_\mu A^\mu (|a|^2 + |b|^2) \nonumber \\ \nonumber \\
&&-\frac{1}{4}F^{\mu \nu}F_{\mu \nu} - 
\frac{1}{2}\bar{\lambda} \not \! \partial \lambda + \frac{1}{2} D^2. 
\eea
The electron is represented by $\psi$ and its propagator is of the general form
\begin{equation} \label{eq:fermiprop}
S(p) \equiv \frac{-i}{\slash p A(p^2) + B(p^2)}
=-i\frac{{\cal Z}(p^2)}{\slash p + {\cal M}(p^2)},
\end{equation}
where ${\cal Z}(p^2), {\cal M}(p^2), \Ap2$ and $\Bp2$ are scalar functions. It 
makes up the chiral multiplet
together with the selectrons $a,f,b,g$. The selectron propagators are 
restricted by SUSY Ward identities~\cite{IZ74,WB99b} to be 
\bea
D_{aa}(p^2)&\equiv & \langle a^* a \rangle (p^2)
= \frac{A(p^2)}{p^2 A^2 (p^2) - B^2 (p^2)}, \\
D_{af}(p^2)&\equiv & \langle a^* f \rangle (p^2)
= \frac{B(p^2)}{p^2 A^2 (p^2) - B^2 (p^2)}, \\
D_{ff}(p^2)&\equiv & \langle f^* f \rangle (p^2)
= \frac{p^2 A(p^2)}{p^2 A^2 (p^2) - B^2 (p^2)}. \label{eq:ffprop}
\eea
The photon and photino fields are $A_\mu$ and $\lambda$, respectively.

The DSE in SQED is given by~\cite{WB99b}
\bea \label{eDSE}
\lefteqn{S^{-1}(p) - S_0^{-1}(p)} \nonumber \\
&=&-\int \frac{d^d q}{(2\pi)^4}\{ D_{\mu \nu}(p-q) \gamma^\mu S(q) 
\Gamma_{\psibar A_\mu \psi}^\nu (q,p) +
S_\lambda (p-q) D_{aa}(q) \Gamma_{\lambdabar a^* \psi}(q,p) \nonumber \\
&&\hspace{1cm} + S_\lambda (p-q) D_{af}(q) \Gamma_{\lambdabar f^* \psi}(q,p)\},
\eea
where the dimensionality $d$ can be either 3 or 4, $S_0$ denotes the bare
propagator and the photon and photino propagators are denoted by 
$D_{\mu \nu}(p-q)$ and $S_\lambda (p-q)$ respectively. The three-point vertices
in Eq.~(\ref{eDSE}) are given by~\cite{WB99b}
\bea \label{thirtysix} 
\Gamma_{\lambdabar a^* \psi}(p,q) &=& \frac{e}{p^2 - q^2}(p^2 \Ap2 - q^2 \Aq2)
+ \frac{e}{p^2 - q^2}(\Bp2 - \Bq2)\slash q \nonumber \\
&& + \half e (p^2 - \slash q \slash p)T_{aa}(p^2,q^2,p\cdot q) \nonumber \\
&& + \half e p^2(q^2 - \slash p \slash q)T_{ff}(p^2,q^2,p\cdot q) 
\nonumber \\
&&+ \half e [\slash p (p^2 - q^2) - 2\slash q (p^2 - p\cdot q)]
T_{af}(p^2,q^2,p\cdot q), 
\eea
\bea \label{thirtyfive} 
\Gamma_{\lambdabar f^* \psi}(p,q) &=& \frac{-e}{p^2 - q^2}(\Ap2 - \Aq2)\slash q
-\frac{e}{p^2 - q^2}(\Bp2 - \Bq2) \nonumber \\
&& +\half e(\slash p - \slash q)T_{aa}(p^2,q^2,p\cdot q)\nonumber \\
&& +  \half e(p-q)^2 T_{af}(p^2,q^2,p\cdot q) \nonumber \\
&& - \half e\slash q(p^2 - \slash p \slash q)T_{ff}(p^2,q^2,p\cdot q),
\eea

and
\bea
\Gamma^\mu_{\psibar A_\mu \psi}(p,q) &=&
\Gamma^\mu_{BC}(p,q) + \frac{ie}{p^2 - q^2}(\Ap2 - \Aq2)
	[\half T_3^\mu - T_8^\mu] \nonumber \\
&&		- \frac{ie}{p^2 - q^2}(\Bp2 - \Bq2)T_5^\mu
		+ \half ie T_{aa}(p^2,q^2,p\cdot q) T_3^\mu \nonumber \\
&& +ie T_{af}(p^2,q^2,p\cdot q)
	[\half (p-q)^2 T_5^\mu - T_1^\mu] \nonumber \\
&& + \half ie T_{ff}(p^2,q^2,p\cdot q)
	[T_2^\mu - p\cdot q T_3^\mu - (p-q)^2 T_8^\mu],
\eea
where
\bea
\Gamma^\mu_{BC}(p,q) &=& \half \frac{ie}{p^2 - q^2}(\slash p + \slash q)
(A(p^2) - A(q^2))(p+q)^\mu \nonumber \\
&& + ie\half (A(p^2) + A(q^2)) \gamma^\mu
+ \frac{ie}{p^2 - q^2}(\Bp2 - \Bq2)(p+q)^\mu,
\eea
\bea \label{transverse}
T_1^\mu &=& p^\mu(q^2 - p\cdot q) + q^\mu (p^2 - p\cdot q), \\
T_2^\mu &=& (\slash p + \slash q)T_1^\mu, \\
T_3^\mu &=& \gamma^\mu(p-q)^2 - (\slash p - \slash q)(p-q)^\mu], \\
T_5^\mu &=& \sigma^{\mu \nu}(p-q)_\nu, \\
T_8^\mu &=& \half(\slash p \slash q \gamma^\mu - \gamma^\mu \slash q \slash p),
\eea
and represent the electron-$a$-photino, 
electron-$f$-photino and electron-photon vertices respectively.

Performing the Wick rotation into Euclidean space 
and substituting the full vertices into the DSE
gives us the following integral equations:
\bea \label{eq:Bfull}
B(p^2) &=& 2e^2\int \frac{d^dq}{(2\pi)^d} \inverse{(p-q)^2}
\inverse{p^2 - q^2}[D_{af}(q^2)p^2 \Ap2 \nonumber \\
&& \hspace{4cm} + (p^2 - 2q^2)D_{aa}(q^2)\Bp2] \nonumber \\
&&+(\xi-1)e^2\int \frac{d^dq}{(2\pi)^d} \inverse{(p-q)^4}
[D_{af}(q^2)p^2 \Ap2 + D_{aa}(q^2)q^2 \Bp2] \nonumber \\
&&-(\xi-1)e^2\int \frac{d^dq}{(2\pi)^d} \frac{p\cdot q}{(p-q)^4}
[D_{af}(q^2) \Ap2 + D_{aa}(q^2) \Bp2] \nonumber\\
&&-\half e^2\int \frac{d^dq}{(2\pi)^d} D_{af}(q^2)
T_{aa}(p^2,q^2,p\cdot q) \nonumber \\
&&-e^2 \int \frac{d^dq}{(2\pi)^d} D_{aa}(q^2) T_{af}(p^2,q^2,p\cdot q)
\left[\frac{(p\cdot q)^2 - p^2 q^2}{(p-q)^2} + q^2 - p\cdot q \right] 
\nonumber \\
&&+\half e^2\int \frac{d^dq}{(2\pi)^d} D_{af}(q^2) 
T_{ff}(p^2,q^2,p\cdot q) 
\left[p^2 \frac{q^2 - p\cdot q}{(p-q)^2} + q^2 \frac{p^2 - p\cdot q}{(p-q)^2}
\right],
\eea
\bea \label{eq:Afull}
\Ap2 - 1 &=& 2e^2\int \frac{d^dq}{(2\pi)^d} \inverse{(p-q)^2}
\inverse{p^2 - q^2}D_{aa}(q^2)\left[(p^2 - 2q^2)\Ap2 + q^2 \Aq2 \right] 
\nonumber \\
&&-2e^2\int \frac{d^dq}{(2\pi)^d} \inverse{(p-q)^2}
\inverse{p^2 - q^2}D_{af}(q^2)[\Bp2 - \Bq2] \nonumber \\
&&+ (\xi -1) e^2\int \frac{d^dq}{(2\pi)^d} \inverse{(p-q)^4}
D_{aa}(q^2) q^2 [\Ap2 + \Aq2 ] \nonumber \\
&& -(\xi -1) e^2\int \frac{d^dq}{(2\pi)^d} \inverse{(p-q)^4}
D_{af}(q^2)[\Bp2 - \Bq2] \nonumber \\
&&+(\xi-1)\frac{e^2}{p^2} \int \frac{d^dq}{(2\pi)^d} 
\frac{p\cdot q}{(p-q)^4}D_{af}(q^2)[\Bp2 - \Bq2] \nonumber\\
&& -(\xi-1)\frac{e^2}{p^2} \int \frac{d^dq}{(2\pi)^d} 
\frac{p\cdot q}{(p-q)^4}D_{aa}(q^2)[p^2 \Ap2 + q^2 \Aq2] \nonumber \\
&&+\frac{e^2}{p^2}\int \frac{d^dq}{(2\pi)^d} 
D_{aa}(q^2) T_{aa}(p^2,q^2,p\cdot q) \nonumber \\
&& \hspace{1.5cm} \left[
\frac{3}{2}p\cdot q - p^2 \frac{q^2 - p\cdot q}{(p-q)^2}
- p\cdot q \frac{p^2 - p\cdot q}{(p-q)^2} \right] \nonumber \\
&&+\half \frac{e^2}{p^2}\int \frac{d^dq}{(2\pi)^d} 
D_{af}(q^2)T_{af}(p^2,q^2,p\cdot q) \nonumber \\
&& \hspace{1.5cm} \left[p\cdot q - 3p^2 
+2(p^2 - p\cdot q)\frac{q^2 - p\cdot q}{(p-q)^2}\right]
 \nonumber \\
&&+\frac{3}{2}\frac{e^2}{p^2}\int \frac{d^dq}{(2\pi)^d} 
D_{aa}(q^2) T_{ff}(p^2,q^2,p\cdot q) q^2 p^2 ,
\eea
where the scalar functions $T_{aa}(p^2,q^2,p\cdot q),T_{af}(p^2,q^2,p\cdot q),
T_{ff}(p^2,q^2,p\cdot q)$ are symmetric in $p,q$ due to charge conjugation
invariance~\cite{WB99b} but are unconstrained by either the Ward-Takahashi 
or SUSY Ward identities. In a slight abuse of language 
we refer to these functions as `transverse' functions since they 
contribute only to the transverse components of the vertices. 
The contribution to $\Bp2$ from the transverse components of the
vertices is referred to as the transverse contribution. 
Eqs.~(\ref{eq:Bfull},\ref{eq:Afull}) were derived in the quenched 
approximation. The effects of vacuum polarisation will be considered shortly.

Considering first the minimal SUSY Ball-Chiu 
{\it ansatz} where the transverse functions
are set to zero, we see immediately that there is
no reason why $\Bp2$ must vanish. We can also eliminate the possibility of 
the transverse functions inducing a nonrenormalisation theorem by
cancelling off the minimal contribution. To see this, recall that
the transverse functions are symmetric in $p$ and $q$ and examine
the coefficients of $D_{af}(q^2)$. Its coefficients
in the minimal contribution are asymmetric in $p,q$ because of the $\Ap2$
factor whereas those in the transverse contribution are exactly symmetric.
(A corresponding argument for terms in Eq.~(\ref{eq:Bfull})
proportional to $D_{aa}(q^2)$ cannot be made because the transverse
contribution has an asymmetric component.) It follows that the integrand of 
$\Bp2$ will not vanish, regardless of the choice of transverse functions. This
result will still hold after angular integration as the transverse contribution
will be a symmetric function multiplied by $q^{d-1}$ whereas the minimal
contribution will not. However we cannot eliminate the possibility that 
Eqs.~(\ref{eq:Bfull},\ref{eq:Afull}) may simply not be solvable unless
$\Bp2$ is set to zero.

This apparent contradiction between ourselves and previous superfield 
analyses~\cite{CL88,CS99a} requires explanation. If an achiral solution is
forbidden in the superfield formalism then it must also be forbidden in the
component formalism, and yet our analysis finds no evidence that an 
achiral solution cannot exist. We can eliminate the possibility that 
our choice of the
quenched approximation has obscured a nonrenormalisation theorem. Indeed, the 
photon propagator is of the general form
\begin{equation} \label{eq:photonprop}
D_{\mu \nu}(p) \equiv \inverse{p^2}(g_{\mu \nu} -\frac{p_\mu p_\nu}{p^2})
\inverse{1 + \Pi(p^2)} + \xi \frac{p_\mu p_\nu}{p^4},
\ee
where $\Pi(p^2)$ is the vacuum polarisation. The photino and
$D$ propagators are then restricted by SUSY to be~\cite{KS89}
\bea \label{eq:lambdaprop}
S_{\lambda}(p) &=& \frac{-i}{\slash p}\inverse{1 + \Pi(p^2)}, \\
D_{D} (p^2) &=& \inverse{p^2} \inverse{1 + \Pi(p^2)} \label{eq:Dprop},
\eea
respectively. In the absence of approximations we can confidently choose to
work in a specific gauge, and in Landau gauge ($\xi=0$) the right hand sides
of Eqs.~(\ref{eq:Bfull},\ref{eq:Afull}) are simply multiplied by a factor of
$\inverse{1 + \Pi(p^2)}$, which preserves our result.

It would seem that the 
vanishing gauge dependent prefactor in superspace treatments~\cite{CL88,CS99a} 
is an artifact of the extensive approximations used. The approximations 
in~\cite{CS99a} were 
generally chosen so as to have minimal impact in the infrared region
where chiral symmetry breaking is largely determined, but to combine such
approximations with a gauge dependent argument is dangerous. Consider, for
example, the
equivalent of Eq.~(\ref{eq:Bfull}) in non-SUSY QED$_3$ in the quenched
rainbow approximation,
\be \label{eq:nonsusyB}
\Bp2 = (\xi+2)\frac{e^2}{4\pi^2 p}\int_0^\infty dq
\frac{q\Bq2}{q^2 A^2(q^2) + B^2(q^2)}(\mbox{ln}|p+q| - \mbox{ln}|p-q|).
\ee
In the special gauge of $\xi=-2$ the right hand side of Eq.~(\ref{eq:nonsusyB})
vanishes unless $\Aq2$ and $\Bq2$ conspire to cancel this prefactor. It does
not follow though that chiral symmetry is unbroken. In fact non-SUSY QED$_3$
is known to break chiral symmetry from lattice studies~\cite{DKK90,HK90}. 
The vanishing prefactor in Eq.~(\ref{eq:nonsusyB}) is 
an artifact of the rainbow approximation.

\setcounter{equation}{0}
\section{THE CJT EFFECTIVE POTENTIAL IN THE NONPERTURBATIVE REGIME}
\label{sec:CJT}

Finding the CJT effective potential for any theory obviously requires the
correct form of the propagators. The correct form of the selectron propagators 
in SUSY theories is made a little obscure by the auxiliary fields $f,g$ and
was obtainable only with the recent realisation~\cite{WB99b} 
that the free Lagrangian
becomes quadratic when the selectrons $a,f$ ($b,g$) form a
column matrix $[a] \equiv \left( \ba{c} a \\ f \ea \right) \hspace{1.5em}
\left([b] \equiv \left( \ba{c} b \\ g \ea \right)\right)$. 
The selectron propagator is 
\be
[D(p^2)] \equiv
\left[ \ba{cc} D_{aa}(p^2) & D_{af}(p^2) \\ 
D_{fa}(p^2) & D_{ff}(p^2) \ea \right]
= \left[ \ba{cc} D_{bb}(p^2) & D_{bg}(p^2) \\ 
D_{gb}(p^2) & D_{gg}(p^2) \ea \right].
\ee

The CJT effective potential is
\bea \label{eq:effpotsqed}
V[S,[D]_a,[D]_b] &=& \int \frac{d^dp}{(2\pi)^d}(\Tr \ln (S^{-1}_0(p) S(p)) +
\half \Tr (1 - S^{-1}_0(p) S(p))) \nonumber \\
&& - 2\int \frac{d^dp}{(2\pi)^d}(\Tr \ln ([D(p^2)]^{-1}_0 [D(p^2)]) 
\nonumber \\ && \hspace{1cm} +
\half \Tr (1 - [D(p^2)]^{-1}_0 [D(p^2)])) \nonumber \\
&& + \half \int \frac{d^dp}{(2\pi)^d}(\Tr \ln (S^{-1}_{0\lambda}(p) 
S_\lambda (p)) + \Tr (1 - S^{-1}_{0\lambda}(p) S_\lambda (p))) \nonumber \\
&& - \half \int \frac{d^dp}{(2\pi)^d}(\Tr \ln (D_{\mu \nu 0}^{-1}(p) 
D_{\nu \sigma}(p)) + 4 - D_{\mu \nu 0}^{-1}(p)D_{\nu \mu}(p)) \nonumber \\
&& - \half \int \frac{d^dp}{(2\pi)^d}(\Tr \ln (D_{0 D}^{-1}(p^2) D_D(p^2))
+ \Tr (1 - D_{0 D}^{-1}(p) D_D(p^2))).
\eea

(The calculation in~\cite{WB99a} used ordinary scalar propagators 
$D_{aa}(p^2)$.) Substituting Eqs.~(\ref{eq:fermiprop}) to (\ref{eq:ffprop}) and
Eqs.~(\ref{eq:photonprop}) to (\ref{eq:Dprop}) into 
Eq.~(\ref{eq:effpotsqed}) reveals 
that the CJT effective potential is zero at its extrema, and therefore
uniformly zero. Note that there is another term in Eq.~(\ref{eq:effpotsqed}), 
given by
\begin{equation} \label{tadpole}
\int \frac{d^3p}{(2\pi)^3} \frac{d^3q}{(2\pi)^3} D_{\mu \nu}(q) 
\Tr([\Gamma_{a^* A_\mu A_\nu a}^{\mu \nu}(p,p,q)][D(p^2)]),
\end{equation}
where $[\Gamma_{a^* A_\mu A_\nu a}^{\mu \nu}]$ is the four-point vertex for two
$[a]$s and two photons, corresponding to the vacuum graphs that
give rise to the tadpole contributions in the boson DSE. As we have
written it, Eq.~(\ref{eq:effpotsqed}) only gives half of these particular
diagrams. However, Eq.~(\ref{tadpole}) is proportional to the massless
tadpole integral and is therefore zero 
by dimensional regularisation~\cite{IZ80}.

So Pisarski's result is not restricted to
the many flavour limit, ie.~the nonrenormalisation theorem for the effective
potential also applies nonperturbatively. 
It is trivial to extend our result to arbitrary numbers of flavours and it
follows that the CJT effective potential cannot be used to select the 
dynamically favoured propagator in a SUSY theory without SUSY breaking.

\section{Conclusion}

There is no reason to suppose the nonexistence of an  achiral 
solution to the DSE in either SQED$_3$ or SQED$_4$, and 
no nonperturbative nonrenormalisation theorem in either of these theories. The
apparent theorem found in  superfield treatments is an artifact of the 
substantial
approximations required by the form of the DSE in the superfield notation.
It must be admitted, however, that an actual achiral solution has so far eluded
us so we have yet to prove that such a solution actually exists, although there
is no evidence that it doesn't. 

Nor have we addressed the issue of whether it
is the chiral or the achiral solution that is dynamically favoured. We have
demonstrated, however, that the CJT effective potential cannot be used to
determine this in the absence of SUSY breaking since it is uniformly zero,
even in the nonperturbative regime.

\section*{Acknowledgements}
The author would like to thank Dr C.J. Burden for many useful and important 
discussions related to the material in this paper. Thank you also to Dr 
L.J.~Tassie, Prof.~N.E.~Mavromatos, and Dr W.~Woolcock for helpful discussions.

%\bibliography{bibli}
\end{document}